\documentclass[a4paper,11pt]{article}

\usepackage{jinstpub} 

\title{\boldmath Diffraction and Smith-Purcell radiation on the hemispherical bulges in a metal plate}

\author[1]{V.V. Syshchenko,\note{Corresponding author.}}
\author{E.A. Larikova,}
\author{Yu.P. Gladkih}

\affiliation{Belgorod State University,\\Pobedy Street, 85, Belgorod 308015, Russian Federation}

\emailAdd{syshch@yandex.ru}

\abstract{The radiation resulting from the uniform motion of a charged particle near a hemispheric bulge in a metal plane is considered. The description of the radiation process based on the method of images is developed for the case of non-relativistic particle and perfectly conducting target. The spectral-angular and spectral densities of the diffraction radiation on the single bulge (as well as the Smith-Purcell radiation on the periodic string of bulges) are computed. The possibility of application of the developed approach to the case of relativistic incident particle is discussed.
}

\keywords{Interaction of radiation with matter; Beam-line instrumentation}

\arxivnumber{} 

\proceeding{XII$^{\text{th}}$ International Symposium <<Radiation from Relativistic Electrons in Periodic Structures>>\\
  September 18-22, 2017\\
  Hamburg, Germany}

\begin{document}
\maketitle
\flushbottom

\section{Introduction}
\label{sec:intro}

The radiation emitted under charged particle traveling near the boundary of the spatially localized target (without crossing it) is called diffraction radiation (DR), whereas the crossing of the target's boundary produces the transition radiation (TR).

One of the ways to describe these types of radiation is the application of the boundary conditions to the Maxwell equations' solutions for the field of the moving particle in two media. It becomes evident that the boundary conditions could be satisfied only after addition the solution of free Maxwell equations that corresponds to the radiation field, see, e.g. \cite{Ter.Mik}.

The conditions on the boundary between vacuum and ideal conductor could be satisfied in some cases via introduction of one or more fictitious charges along with the real charged particle; this approach to electrostatic problems is known as the method of images, see, e.g., \cite{Jackson}. Namely the method of images had been used in the pioneering paper \cite{Ginz.Frank} where TR on a metal plane had been predicted. The method of images had been used also in \cite{Askaryan} for consideration of TR under passage of the particle through the center of the ideally conducting sphere in dipole approximation.

DR and TR of a charge incident on a perfectly conducting sphere under arbitrary impact parameter had been studied using the method of images in \cite{ShSy}. Here we consider DR on the hemispherical bulge in a perfectly conducting plane using the same approach.

\section{DR on hemispherical bulge}\label{DR.section}

\begin{figure}[htbp]
\centering
\includegraphics[width=\textwidth]{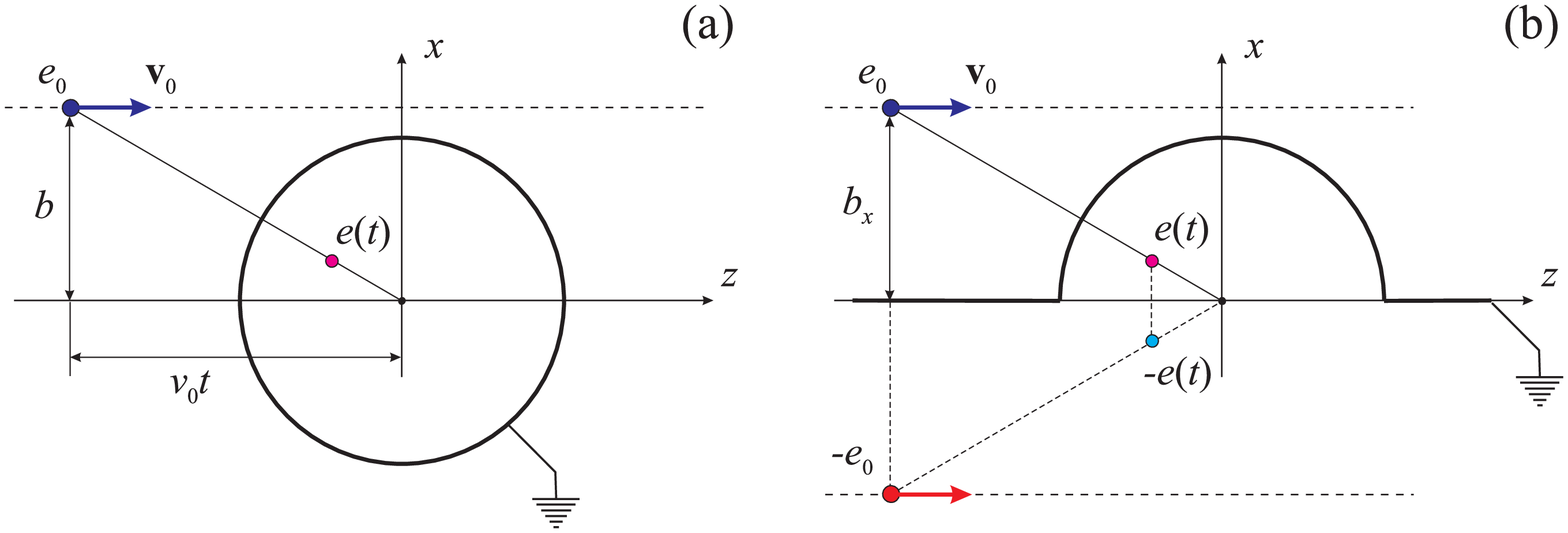} \\
\includegraphics[width=0.65\textwidth]{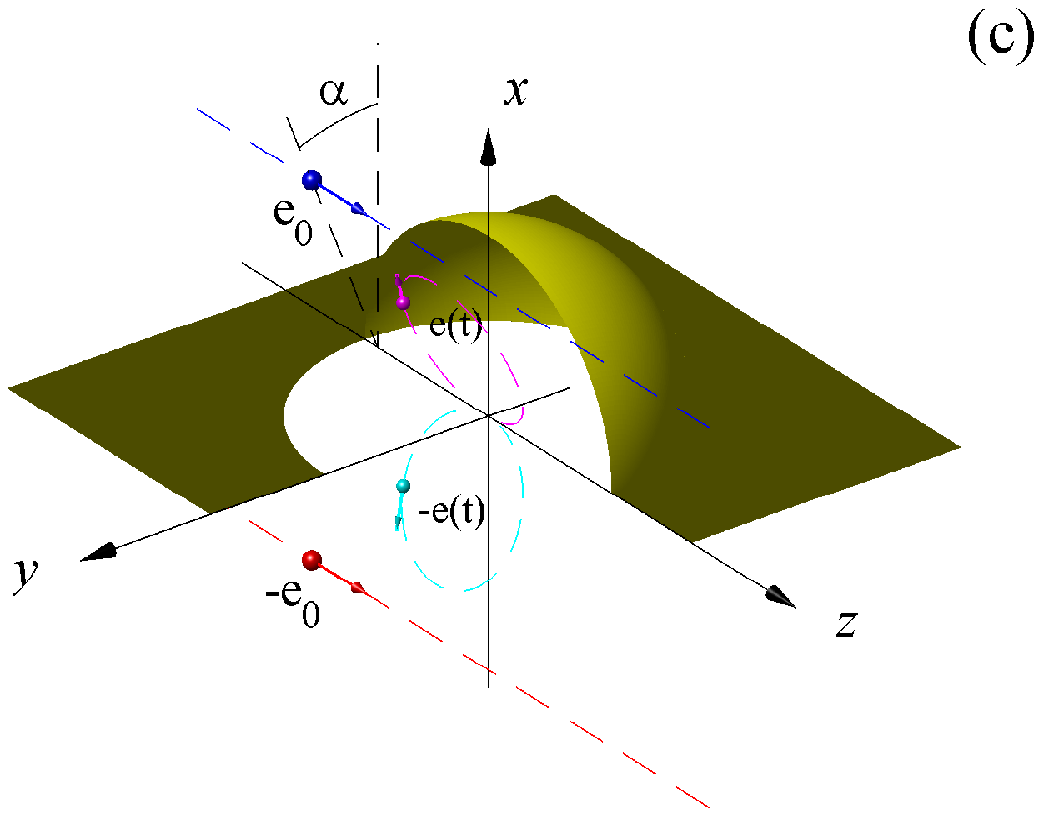} 
\caption{\label{Fig1} The real charge $e_0$ near the grounded sphere and its image (a); the real charge near the hemispherical bulge in the conducting plane and three images (b) and (c).}
\end{figure}

Remember how the method of images is implied to meet the boundary condition for the electric field on the conducting surface. Consider the real charge $e_0$ passing near the grounded conducting sphere of the radius $R$ with the constant velocity $v_0 \ll c$, see fugure \ref{Fig1} (a). In this case the zero potential on the metal surface is acieved via introduction of the single fictitious charge (the ``image'') of the magnitude
\begin{equation}\label{fiktiv.1}
e(t) = -e_0 \frac{R}{\sqrt{b^2 + v_0^2 t^2}}
\end{equation}
placed at the point with coordinates
\begin{equation}\label{fiktiv.2}
x(t) = \frac{R^2b}{b^2 + v_0^2 t^2}, \ z(t) = \frac{R^2 v_0 t}{b^2 + v_0^2 t^2}.
\end{equation}
While the incident particle moves uniformly, its image will move accelerated. 
The radiation produced by non-uniform motion of the image charge will be described by the well-known formula \cite{AhSh}
\begin{equation}\label{spectral.angular.0}
\frac{d\mathcal E}{d\omega d\Omega} = \frac{1}{4\pi^2c} \left|\mathbf k \times \mathbf I^{\vphantom{2}} \right|^2 ,
\end{equation}
where $\mathbf k$ is the wave vector of the radiated wave, $|\mathbf k| = \omega/c$, and
\begin{equation}\label{vector.I}
\mathbf I = \int_{-\infty}^\infty e(t)\,\mathbf v(t)\, \exp\{i(\omega t - \mathbf k\mathbf r(t))\} \, dt
\end{equation}
(it could be easily seen that it is applicable to the case of time-varying charge $e(t)$ as well as to the case of the constant one). Note that the method of images for the isolated (in contrast to grounded) sphere requires introducing another fictitious charge of the magnitude $-e(t)$ resting at the center of the sphere. However, the last equation shows that such rest charge does not produce any radiation.

To achieve the zero potential on the conducting plane with hemispherical bulge we need three image charges along with the real one: the image of the incident charge by the spherical surface (\ref{fiktiv.1})--(\ref{fiktiv.2}) and the mirror reflections of both of them by the plane, $-e_0$ and $-e(t)$, see figure \ref{Fig1} (b). Two of these image charges, $e(t)$ and $-e(t)$, will move accelerated under uniform motion of the incident particle and hence produce the radiation. 

Note, however, that the projectile can fly not only above the top of the hemisphere. For account of this possibility we introduce the two-dimensional impact parameter $\mathbf b = (b_x , b_y) = b\,(\cos\alpha , \sin\alpha )$, see figure \ref{Fig1} (c), so we have
\begin{equation}\label{fiktiv.3}
x^{(1)}(t) = \frac{R^2b_x}{b^2 + v_0^2 t^2}, \ 
y^{(1)}(t) = \frac{R^2b_y}{b^2 + v_0^2 t^2}, \ 
z^{(1)}(t) = \frac{R^2 v_0 t}{b^2 + v_0^2 t^2}
\end{equation}
for the image charge $e(t)$ and
\begin{equation}\label{fiktiv.4}
x^{(2)}(t) = -x^{(1)}(t) , \ 
y^{(2)}(t) = y^{(1)}(t), \ 
z^{(2)}(t) = z^{(1)}(t)
\end{equation}
for the image charge $-e(t)$.

Substituting the proper values into (\ref{vector.I}) and collecting together the contributions from these two image charges, we obtain the following integrals:
\begin{equation}\label{I.x.pres}
I_x = 4e_0 b_x R^3 v_0^2 \int_{-\infty}^\infty 
\exp\left\{i\left[ \omega t - \frac{k_yb_y R^2}{b^2 + v_0^2 t^2} - \frac{k_z R^2 v_0t}{b^2 + v_0^2 t^2}\right] \right\}
\cos \frac{k_xb_x R^2}{b^2 + v_0^2 t^2}
 \, \frac{t \,dt}{(b^2 + v_0^2 t^2)^{5/2}} ,
\end{equation}
\begin{equation}\label{I.y.pres}
I_y = -4ie_0 b_y R^3 v_0^2 \int_{-\infty}^\infty 
\exp\left\{i\left[ \omega t - \frac{k_yb_y R^2}{b^2 + v_0^2 t^2} - \frac{k_z R^2 v_0t}{b^2 + v_0^2 t^2}\right] \right\}
\sin \frac{k_xb_x R^2}{b^2 + v_0^2 t^2} 
\, \frac{t \,dt }{(b^2 + v_0^2 t^2)^{5/2}},
\end{equation}
\begin{equation}\label{I.z.pres}
I_z = 2ie_0 R^3 v_0 \int_{-\infty}^\infty 
\exp\left\{i\left[ \omega t - \frac{k_yb_y R^2}{b^2 + v_0^2 t^2} - \frac{k_z R^2 v_0t}{b^2 + v_0^2 t^2}\right] \right\}
\sin \frac{k_xb_x R^2}{b^2 + v_0^2 t^2}  
\, \frac{(b^2 - v_0^2 t^2) \,dt}{(b^2 + v_0^2 t^2)^{5/2}}.
\end{equation}

\begin{figure}[htbp]
\centering
\includegraphics[width=0.49\textwidth]{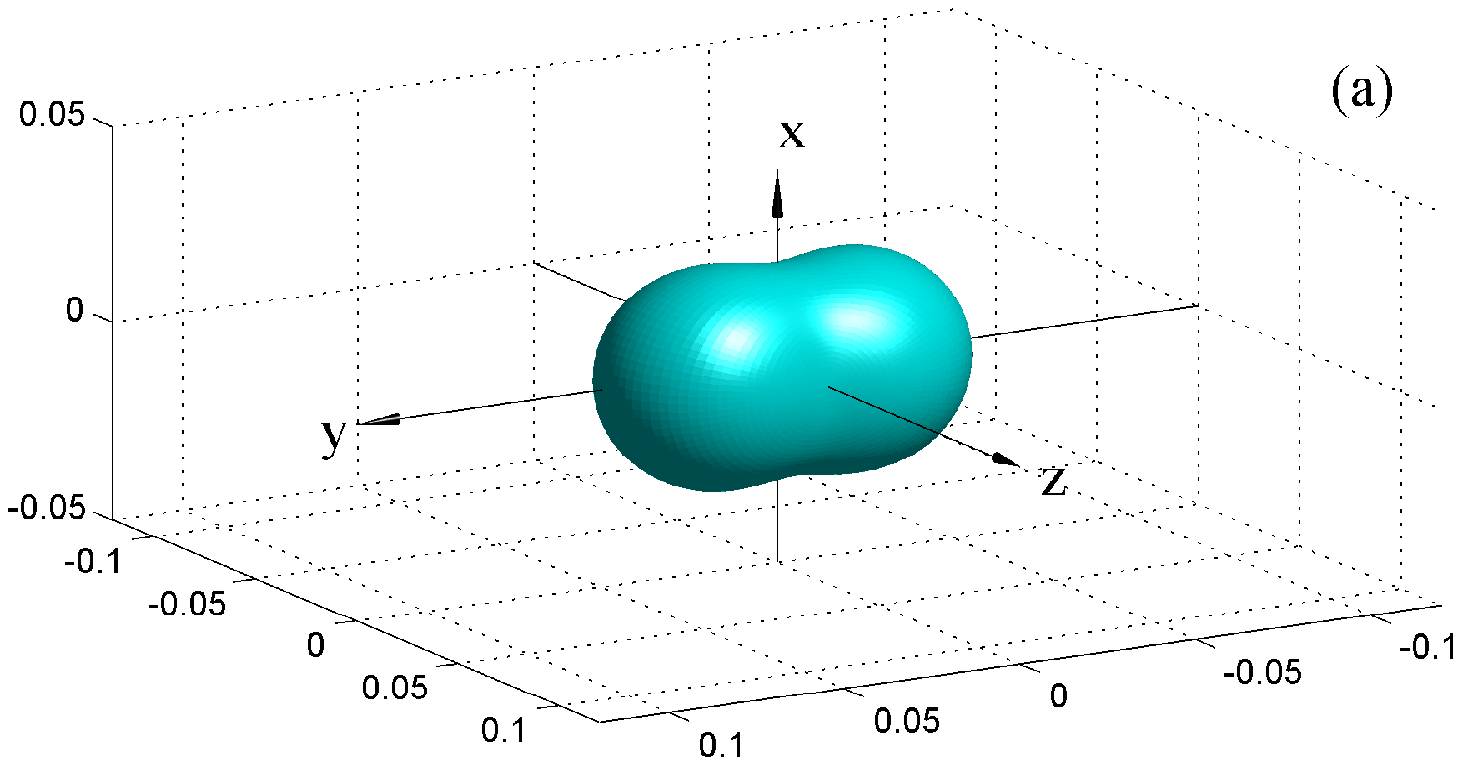} \
\includegraphics[width=0.49\textwidth]{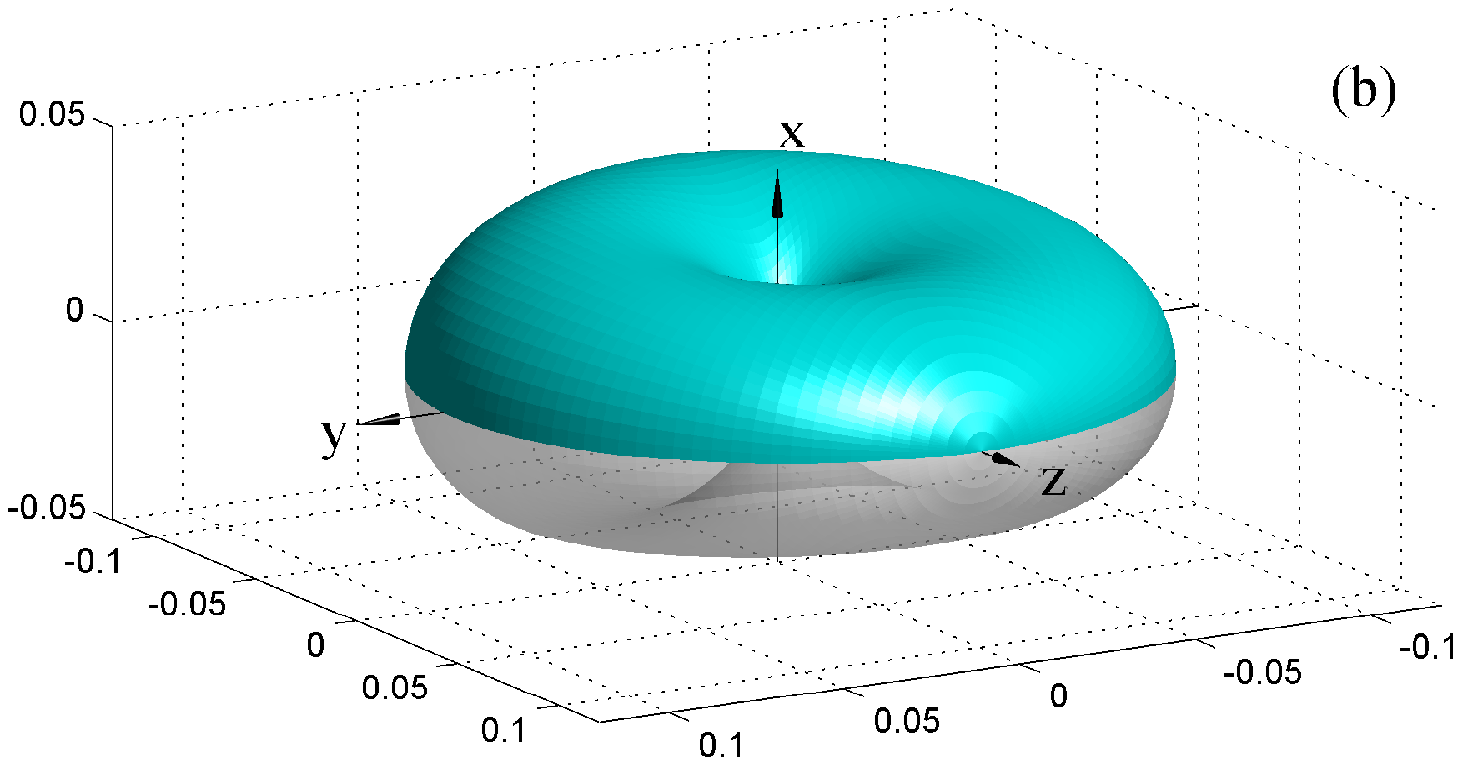} \\
\includegraphics[width=0.49\textwidth]{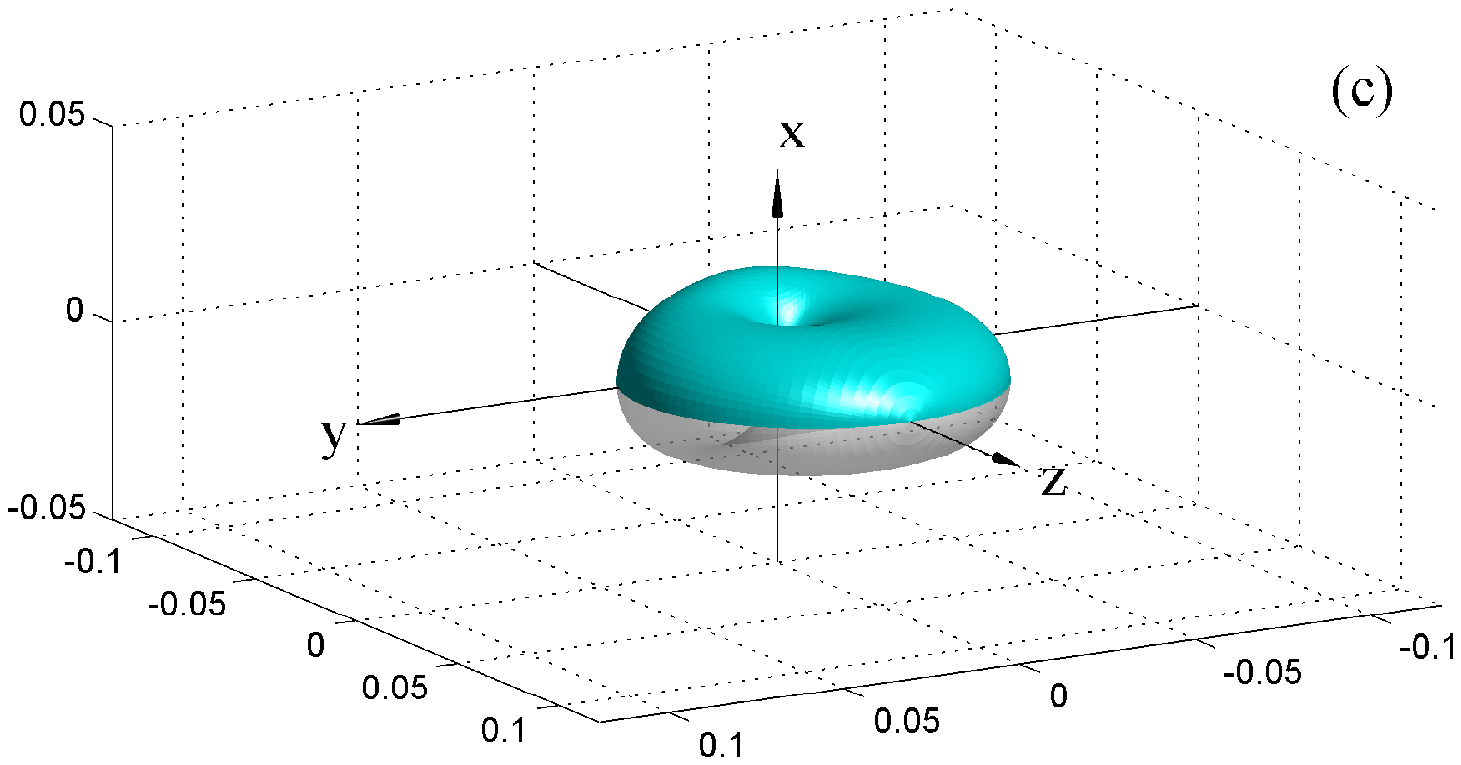} \
\includegraphics[width=0.49\textwidth]{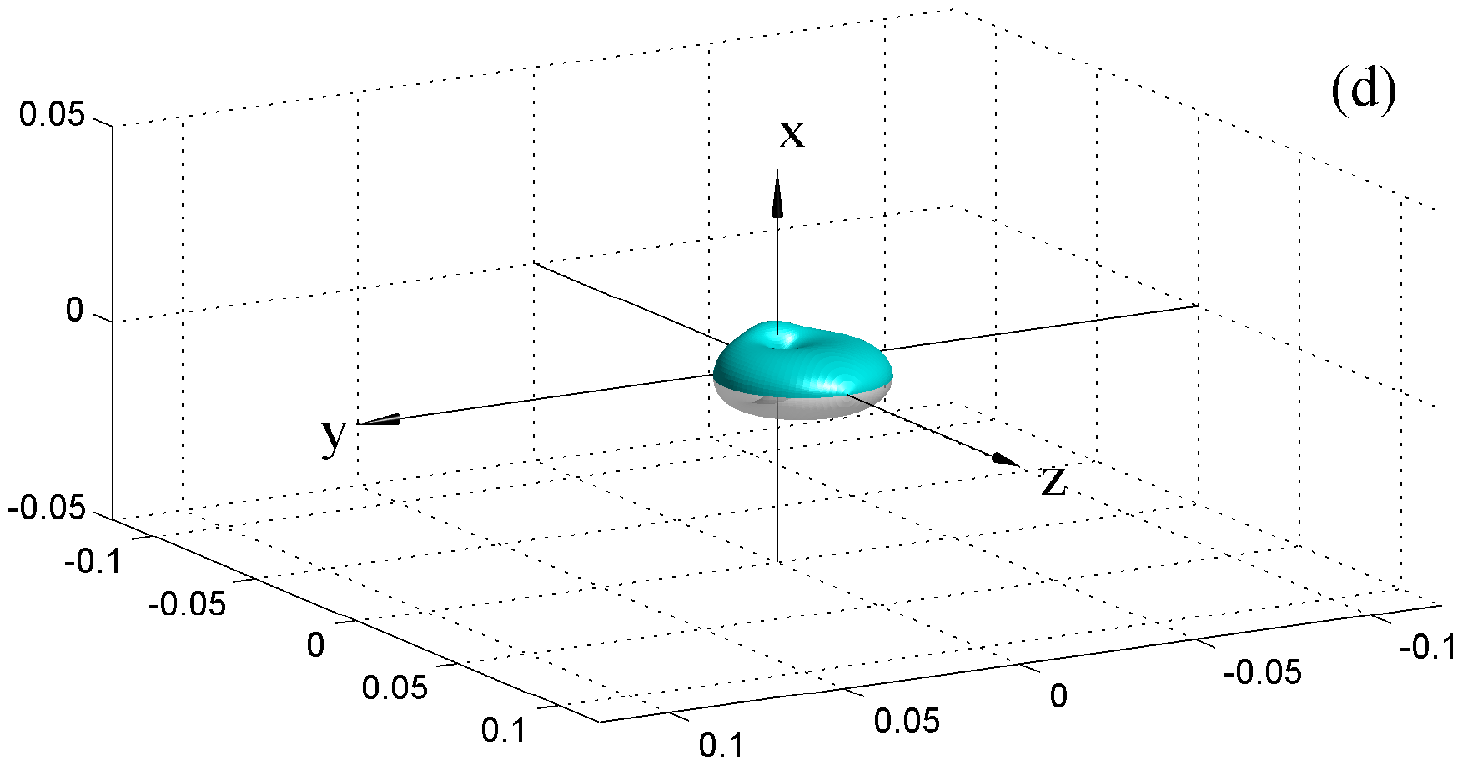}
\caption{\label{Fig2} The angular dependence as direction diagram of DR intensity on the sphere \cite{ShSy} (a) and hemisphere (b) for the passage of the real charge under $b_x = R+0$ and $b_y = 0$ (sliding incidence, when DR intensity is maximal for the whole range of wavelengths) and $R\omega/v_0 = 2.4$ (this choice is due to the maximum of DR on hemisphere spectrum (see below) falls on $\omega b/v_0 \approx 2.4$ and $b = R$ in the given case). This shape of the directional diagram is typical; for higher frequencies the slight forward-backward asymmetry increases, see $R\omega/v_0 = 4$ (c)  and $R\omega/v_0 = 5$ (d).}
\end{figure}

The integrands in (\ref{I.x.pres})--(\ref{I.z.pres}) are smooth functions and the integration can be easily performed numerically, that leads to the spectral-angular density of diffraction radiation in the form
\begin{equation}\label{spectral.angular.1}
  \frac{d\mathcal E}{d\omega d\Omega} = \frac{e_0^2}{4\pi^2c} \, \Phi (\theta , \varphi , \omega) ,
\end{equation}
where the typical shape of the angular distribution $\Phi(\theta , \varphi , \omega)$ is presented in figure \ref{Fig2} (b). We see approximate symmetry of the directional diagram around $x$ axis; for higher $R\omega /v_0$ values the small forward-backward asymmetry increases, see figure \ref{Fig2} (c, d).

\begin{figure}[htbp]
\centering
\includegraphics[width=0.54\textwidth]{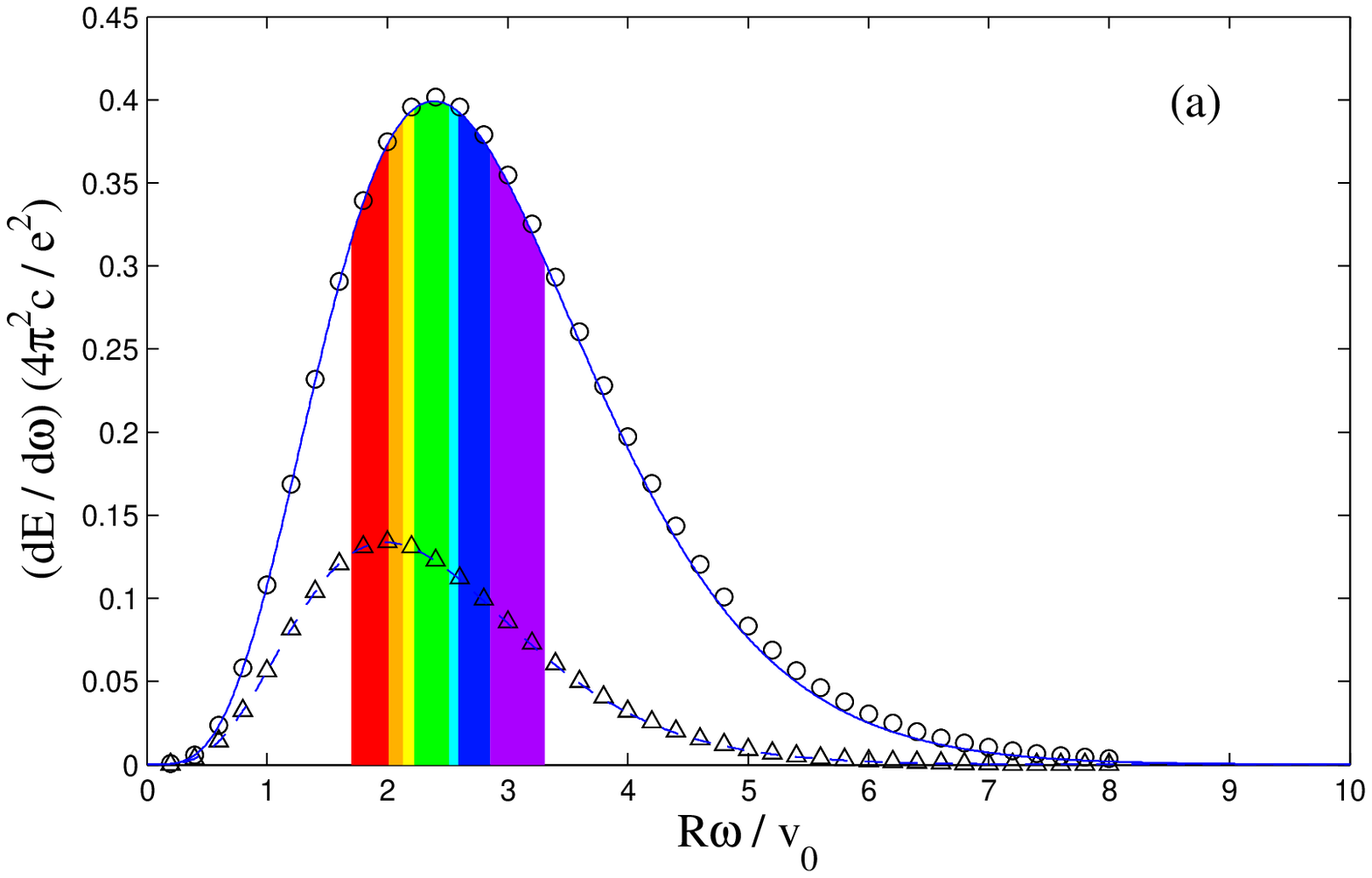} \
\includegraphics[width=0.44\textwidth]{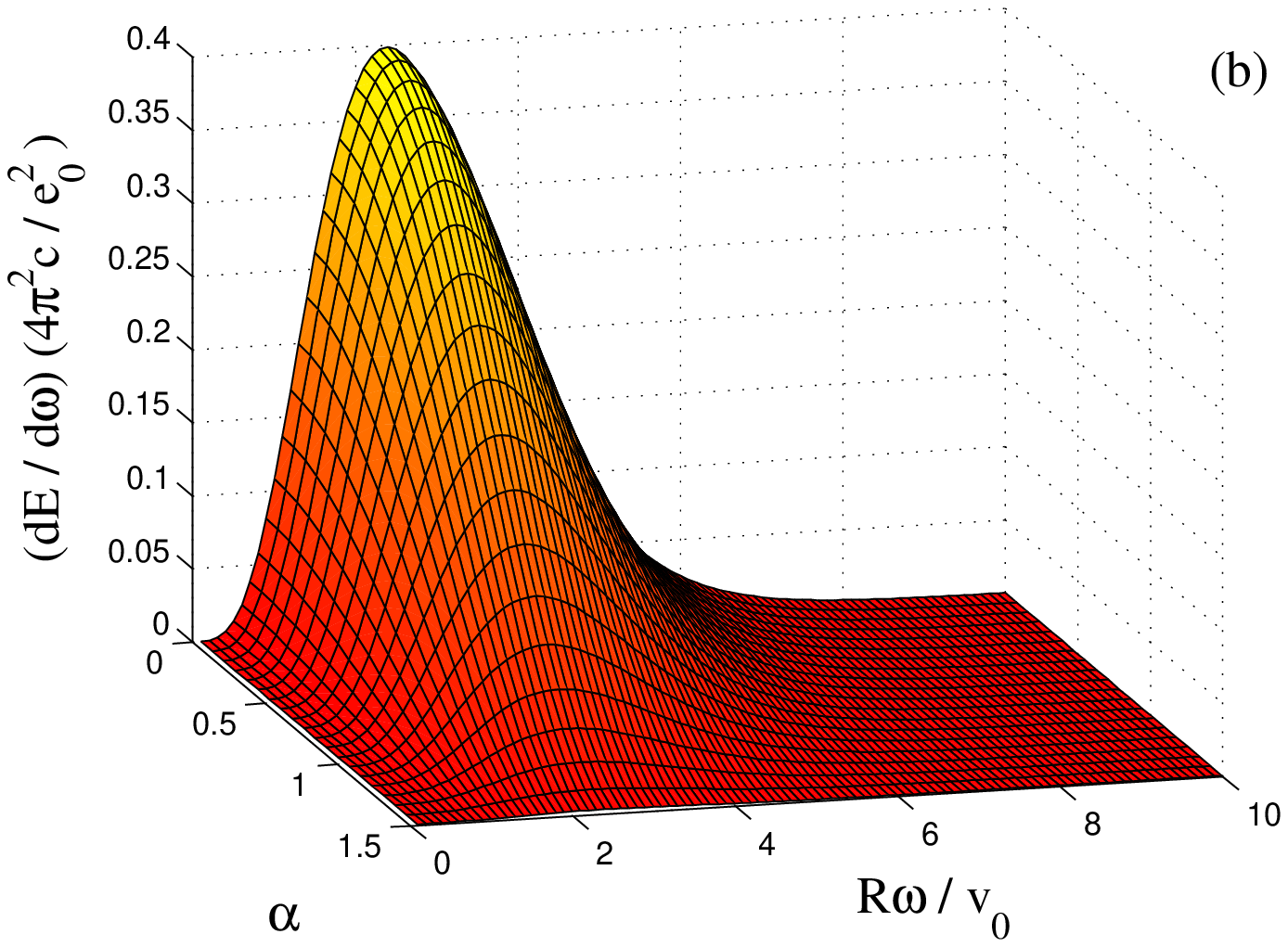} 
\caption{\label{Fig3}(a)  DR spectrum computed using approximated analytical formula (\ref{spectral.angular.2}) (solid curve for $b_ = R+0$, $b_y = 0$ and dashed curve for $b_x = 1.2R$, $b_y=0$)) and via numerical integration (circles and triangles, respectively). (b) For nonzero $b_y$ the radiation intensity decreases approximately as $\cos^2\alpha$ (without substantial changes in the direction diagram shape) .}
\end{figure}

In non-relativistic ($v_0 \ll c$) and low-frequency ($\omega \ll cb/R^2$ hence $\lambda  \gg 2\pi R^2/b$) case we can neglect the second and third terms in the exponents in (\ref{I.x.pres})--(\ref{I.z.pres}) as well as put zero the arguments of the trigonometric functions there. In this case $I_y = I_z = 0$ (that is the reason of the approximate x-axial symmetry of the angular distribution mentioned above) and we obtain the following approximate formula for DR spectral-angular density:
\begin{equation}\label{spectral.angular.2}
\frac{d\mathcal E}{d\omega d\Omega} =  \frac{16}{9\pi^2}  \frac{e_0^2\omega^6 R^6}{c^3 v_0^4} \cos^2\alpha \left( 1 - \sin^2\theta\cos^2\varphi \right)  \left[ K_1 \left(\frac{\omega}{v_0} b \right) \right]^2 ,
\end{equation}
where $K_1$ is the modified Bessel functions of the third kind. This analytical approximation is rather good, as can be seen from the numerical spectrum (integrated (\ref{spectral.angular.1}) over radiation angles) compared with analytical one (figure \ref{Fig3} (a)). 
For illustrative purposes, we choose the parameters $v_0 = 0.1c$, $R = 20$ nm, $b = R+0$, for which the DR intensity maximum will lie in the visible spectrum. The applicability of our results in this frequency domain will be discussed in the Conclusion.

\section{DR on the string of hemispheres}
\label{DR.string.section}
Now consider the motion of the charge $e_0$ along the periodic string of $N\gg 1$ hemispheres. The mutual influence of the fictitious charges induced in the neighboring bulges can be neglected in the case of small impact parameters, $b\to R$ (when the radiation intensity is high), for the string period large enough, $a\gtrsim 5R$. Then the interference of the radiation produced on the subsequent hemispheres leads to the simple formula for the spectral-angular density of DR:
\begin{equation}\label{spectral.angular.string.1}
  \frac{d\mathcal E}{d\omega d\Omega} = \frac{e_0^2}{4\pi^2c} \, \Phi(\theta , \varphi , \omega) 
    \cdot 2\pi N \frac{v_0}{\omega a} \sum_{m=1}^\infty \delta \left(1 - \frac{v_0}{c} \cos\theta - m \frac{2\pi v_0}{\omega a} \right) ,
\end{equation}
where the delta-function means well-known Smith--Purcell condition \cite{Smith.Purcell}.

\begin{figure}[htbp]
\centering
\includegraphics[width=0.49\textwidth]{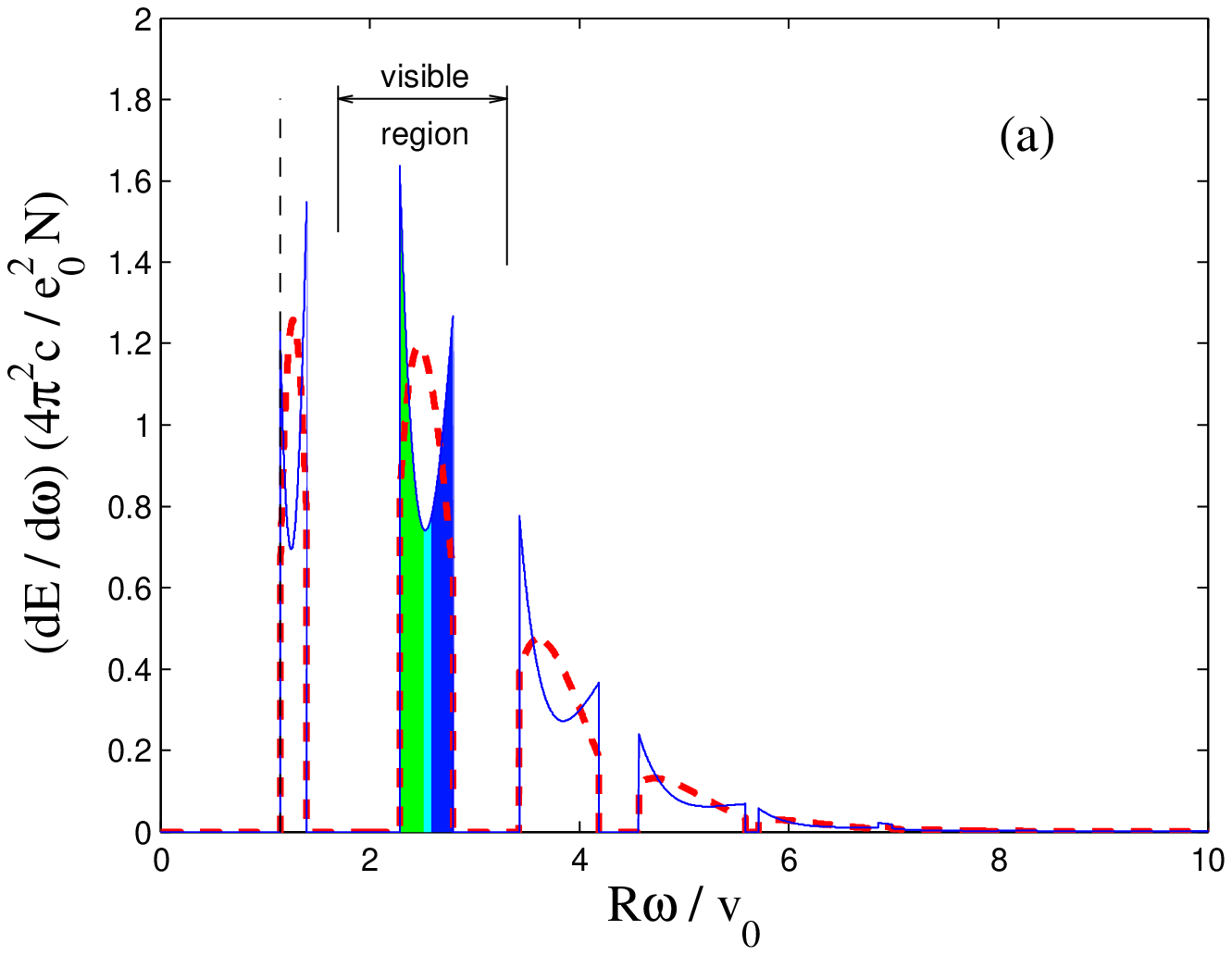} \
\includegraphics[width=0.49\textwidth]{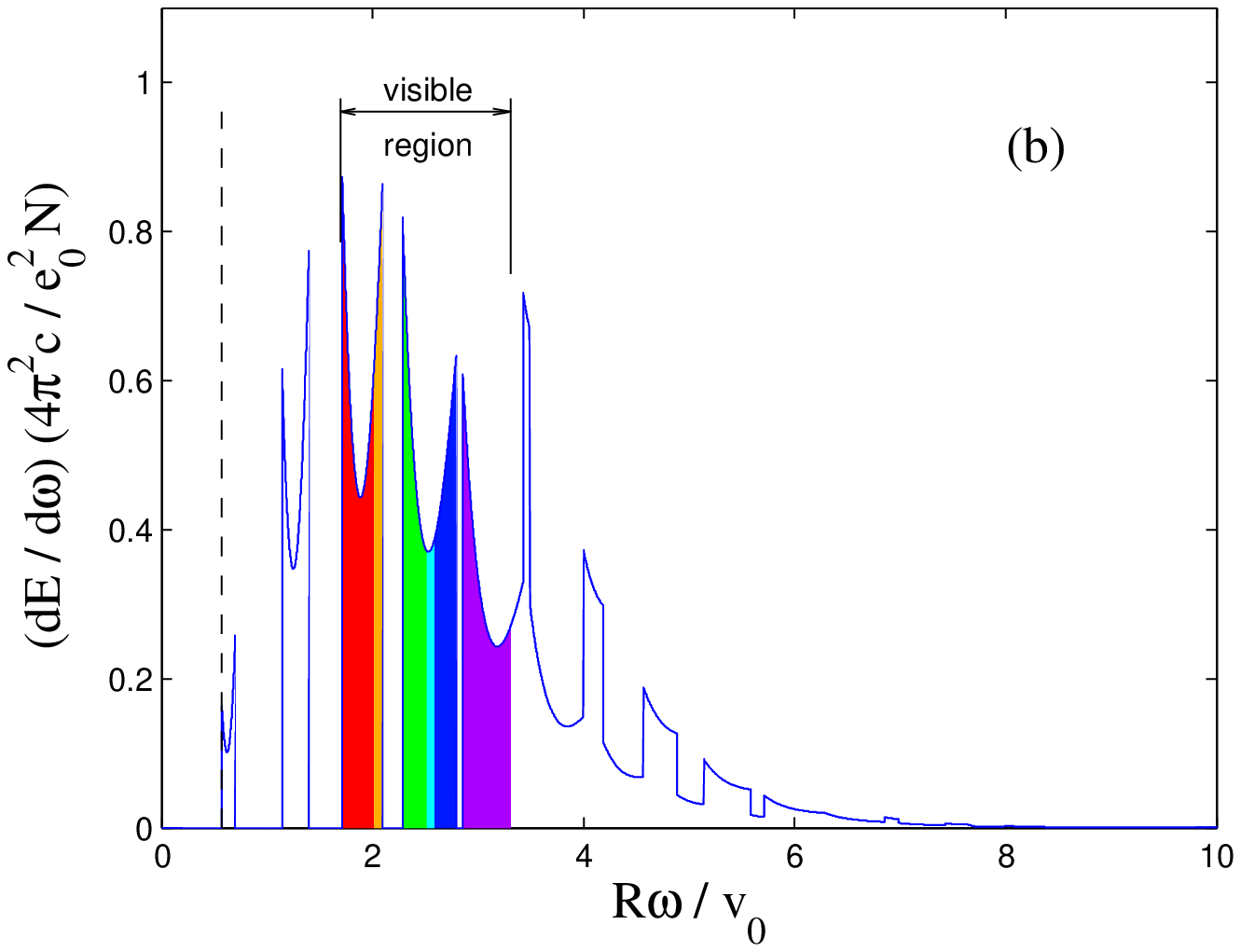} \\
\includegraphics[width=0.49\textwidth]{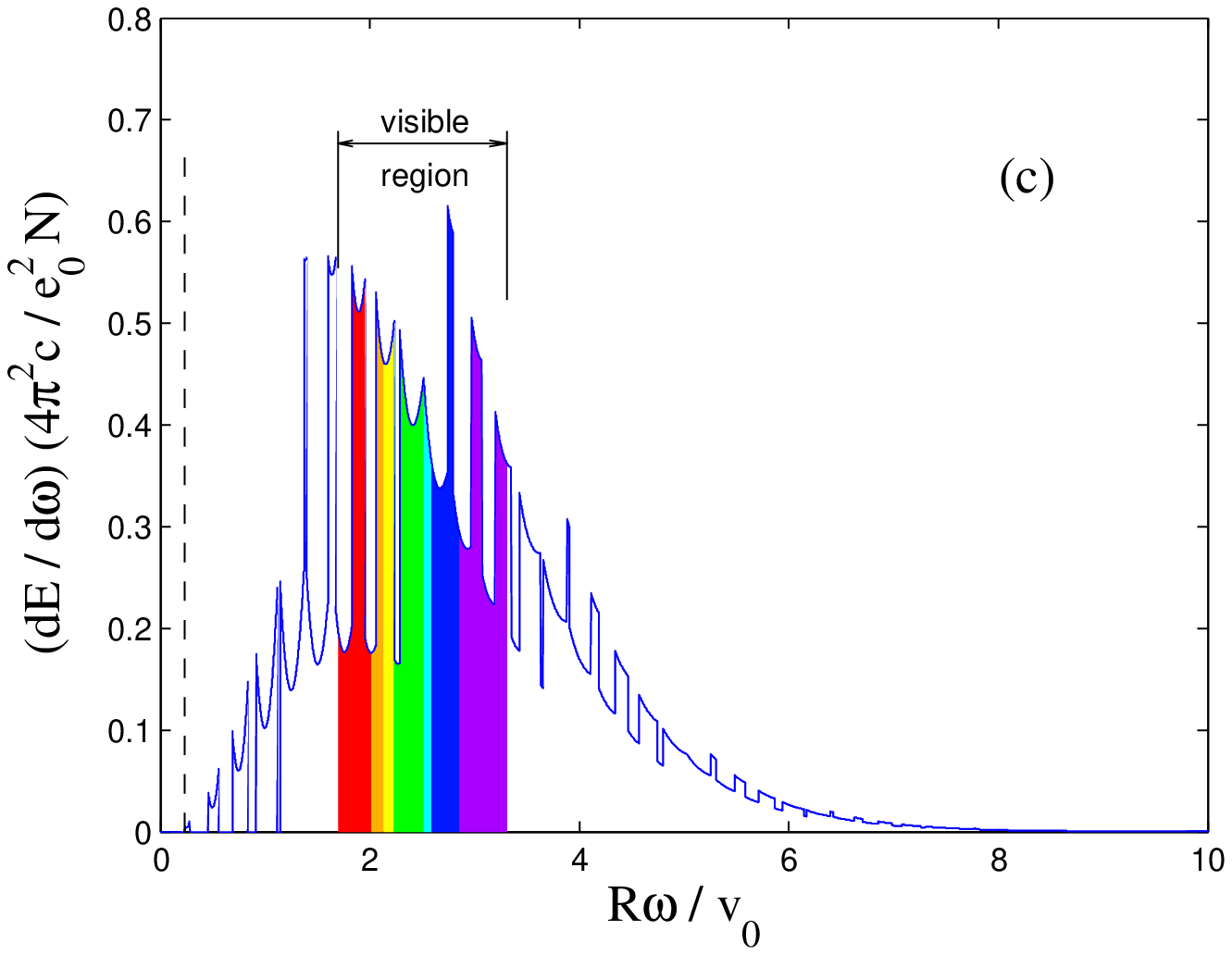} \
\includegraphics[width=0.49\textwidth]{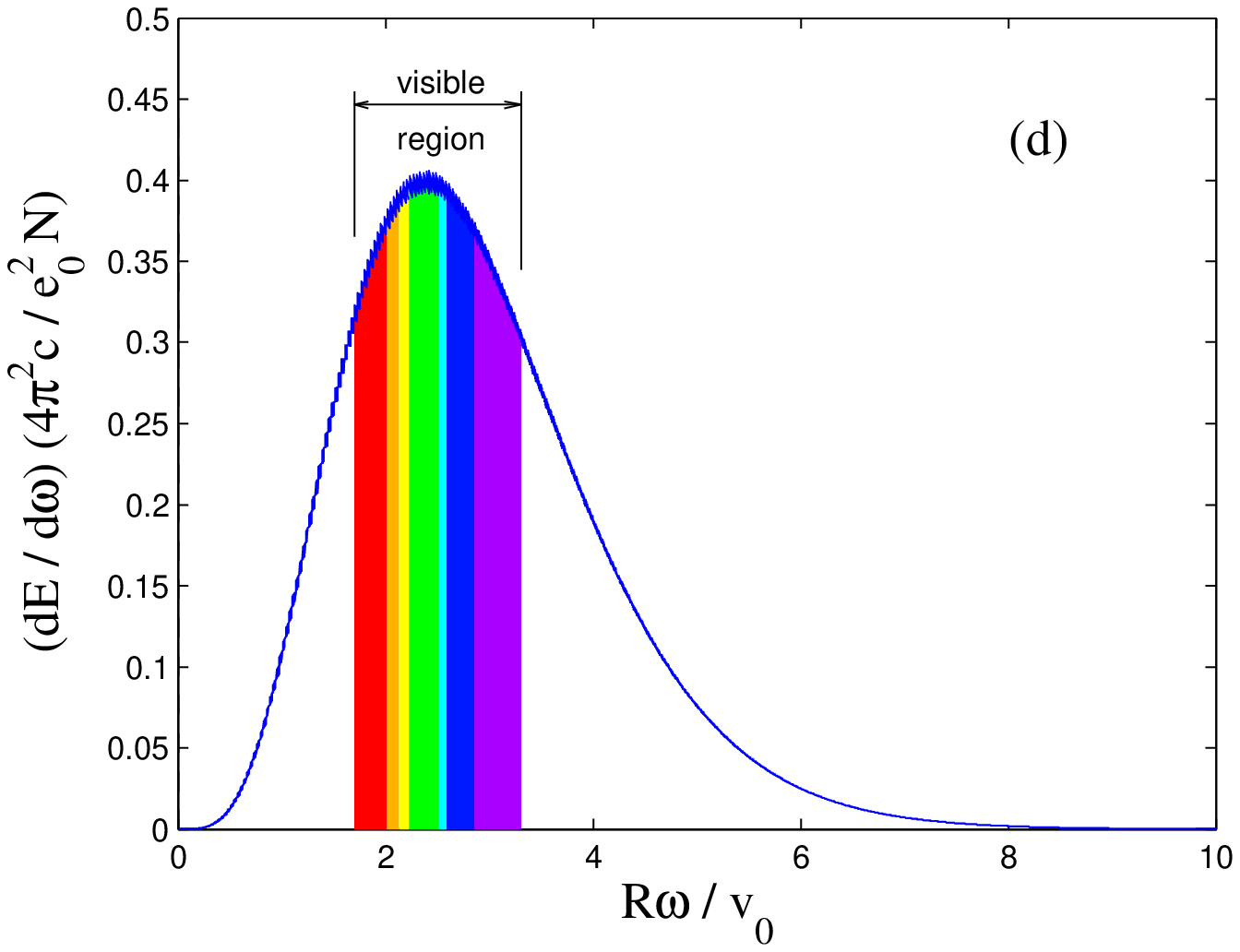}
\caption{\label{Fig4} DR spectrum on the string of spheres under $b = R+0$, $R = 20$ nm, $v_0 = 0.1c$, $a = 100$, 500, 2000, 20000 nm. Vertical dashed line marks the long wavelength edge of the spectrum. Red dotted lines on the panel (a) correspond to DR on the string of spheres \cite{ShSy} under the same conditions.}
\end{figure}

The spectrum for the string period $a$ small enough consists of separated bands, see figure \ref{Fig4} (a). The bands overlap each other under increase of the string period $a$ gradually forming the spectrum of DR on a single hemisphere (multiplied by the total number of the bulges $N$), see figure \ref{Fig4} (b, c, d).

\section{Conclusion}
\label{conclusion}

The diffraction radiation resulting from the interaction of non-relativistic particle with the hemospherical bulge in perfectly conducting plane is considered. The method of images allows the precise description of the radiation in this case. The integration in the resulting formulae can be easily performed numerically that permits to compute the spectral-angular density of the diffraction radiation for an arbitrary impact parameter. The approximate analytical formulae for the radiation characteristics are obtained for the case of radiation wavelengths exceeding the bulge's size. 

The range of validity of our results is determined, first of all, by the validity of the perfect conductor approximation for the metal target that is necessary for the use of the method of images. The perfect conductor approximation means the possibility of the metal's electrons to trace out instantly the changes of the external electric field to meet the requirement of zero tangential component of the electric field on the metal surface. It is valid for the frequencies less than the inverse relaxation time $\tau^{-1}$ for the electrons in the metal. For instance, $\tau^{-1} = 5\cdot 10^{-13}$ sec$^{-1}$ for copper \cite{Kittel}, so the results obtained surely could be applied up to THz and far infrared range. On the other hand, Ginzburg and Tsytovich \cite{Ginz.Tsytovich} wrote: ``\emph{However, a good metal mirror (for instance copper or silver) is in practice fairly close to an ideal mirror for frequencies not higher than the optical range}.'' This means the applicability of the method of images also in the visible domain. However, the surface plasma oscillations also could be important here that needs further investigation.

\begin{figure}[htbp]
\centering
\includegraphics[width=\textwidth]{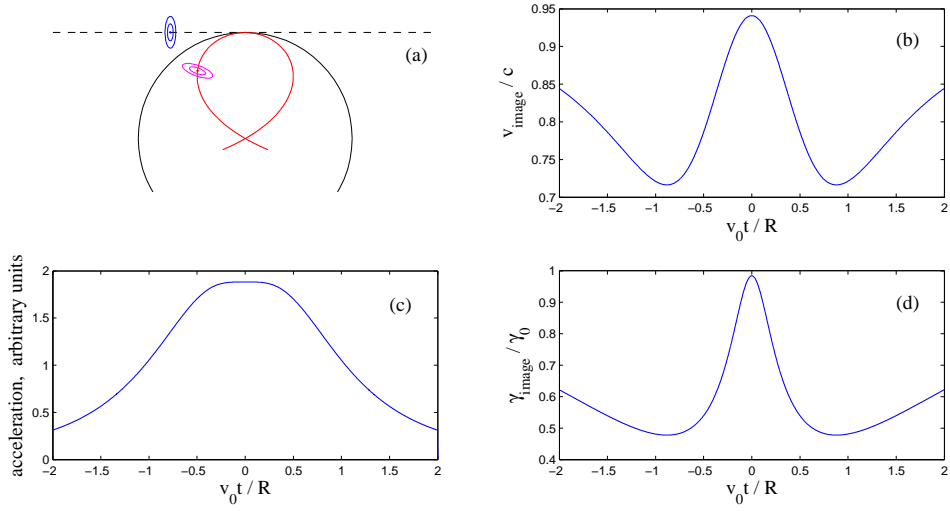} \\
\caption{\label{Fig5} Relativistic incident particle ($\gamma_0 = 3$, $b=1.001R$) and its image in the sphere in locally plane mirror approximation (a) and the characteristics of the image's motion (b, c, d).}
\end{figure}

The results obtained are valid only for the non-relativistic particles due to the geometric nature of the method of images: we can fit the image charge to meet the zero boundary condition on the sphere for the simple Coulomb field of slow incident particle, not for the relativistically compressed one. The only possibility to extend the method-of-images-based approach to relativistic case is to consider the high frequency limit, where the characteristic size of the Coulomb field $v_0\gamma_0/\omega$ (where $\gamma_0 = (1-v_0^2/c^2)^{-1/2}$ is the incident particle's Lorentz factor) is much smaller than the sphere radius $R$. Here we can neglect the metal surface curvature and consider the reflection of the incident field in the locally plane mirror (figure \ref{Fig5}).

However, we have a difficulty also in this case: the kinematic velocity of the image charge (as well as the corresponding Lorenz factor) is not consistent with the degree of the relativistic compression of the Coulomb field of the incident particle. The resulting discrepancy in Lorentz factor values of the real and image charges is negligibly small only for the impact parameters extremely close to the sphere radius, $b = R+0$ and for small part of the incident particle's trajectory. Namely, estimations show that $\gamma_0 - \gamma_{image} \ll \gamma_0$ when
\begin{equation}
    1 - {R}/{\sqrt{b^2 + v_0^2t^2}} \ll \gamma_0^{-2}/4 .
\end{equation}
Due to the constant acceleration of the image charge on the corresponding part of its trajectory (see figure \ref{Fig5} (c)) one could expect synchrotron-like radiation in this case.

\acknowledgments

This research is partially supported by the grant of Russian Science Foundation (project 15-12-10019).

\end{document}